\documentclass[aps,pra,floatfix,twocolumn,lengthcheck,superscriptaddress,showpacs,preprintnumbers,amsmath,amssymb]{revtex4}

\usepackage{graphicx}
\usepackage{dcolumn}
\usepackage{bm}

\begin{document}

\title{Fractional-filling loophole insulator domains for ultracold bosons in optical superlattices}

\author{P. Buonsante}
\affiliation{Dipartimento di Fisica, Politecnico di Torino and I.N.F.M, Corso Duca degli Abruzzi 24 - I-10129 Torino (ITALIA)}%
\author{V. Penna}%
\affiliation{Dipartimento di Fisica, Politecnico di Torino and I.N.F.M, Corso Duca degli Abruzzi 24 - I-10129 Torino (ITALIA)}%
\author{A. Vezzani}
\affiliation{Dipartimento di Fisica, Universit\`a degli Studi di Parma and I.N.F.M., Parco Area delle Scienze 7/a I-43100 Parma (ITALIA)}%

\date{\today}

\begin{abstract}
The zero-temperature phase diagram of a Bose-Einstein condensate
confined in realistic one-dimensional $\ell$-periodic optical superlattices is investigated.
The system of interacting bosons is modeled in terms of a Bose-Hubbard
Hamiltonian whose site-dependent local potentials and hopping amplitudes
reflect the periodicity of the lattice partition in $\ell$-site
cells.
Relying on the exact mapping between the hard-core limit of the
boson Hamiltonian and the model of spinless noninteracting fermions,
incompressible insulator domains are shown to exist  for 
rational fillings that are predicted to be compressible in the atomic
limit. 
The corresponding boundaries, qualitatively described in a  {\it multiple-site} mean-field approach, are shown to exhibit an unusual loophole shape.
A more quantitative description of the loophole domain boundaries at half filling  for the special case $\ell$ = 2 is supplied in terms of  analytic strong-coupling expansions and quantum Monte Carlo simulations. 
\end{abstract}


\maketitle

Ultracold bosons trapped in optical lattices provide a direct realization
of the Bose-Hubbard (BH) model originally introduced to describe liquid He in 
confined geometries \cite{A:Fisher,A:Jaksch}. In this framework, the sites of 
the ambient lattice
correspond to the local minima of the effective potential
produced by the interference pattern of  counterpropagating laser beams. The bosons move across the lattice with a hopping amplitude whose simple relation with the intensity of the laser beams \cite{A:Jaksch}  allows an unprecedented tuning of the model parameters. 
Such control played a key role in a recent breakthrough experiment by Greiner and co-workers \cite{A:Greiner}, who were able to reveal the superfluid-insulator quantum phase transition characterizing the Bose-Hubbard model \cite{A:Fisher}.

The superposition of optical lattices with different periods allows to generate periodic trapping potentials characterized by a richer spatial modulation, the so-called {\it optical superlattices} \cite{A:Peil,A:Guidoni98,A:Roth03,A:Santos,A:LobiMF}. 
As it is well known, in general the single particle spectrum of superlattice (SL) models features band-gaps that critically affect the physical properties of the system. 
In this paper we investigate the zero-temperature phase diagram of the superlattice BH model, showing that classes of parameter choices exist giving rise to unusual loophole-shaped fractional-filling insulator domains. In order to reduce unessential details, we focus on the emblematic case of realistic one-dimensional  superlattices with simple periodicity, that can be created following the scheme described in Ref.~\onlinecite{A:Peil}. In this case, the presence of  insulator loophole domains is directly related to the band-gaps occurring in the single-particle spectrum of the model. The extension of our results to generic superlattices will be presented elsewhere.

The BH Hamiltonian on a 1D superlattice comprising $M$ sites reads
\begin{eqnarray}
\label{E:BHs}
  H = \sum_{k=1}^{M}\!\! &\Big[&\!\!\frac{U}{2} n_k (n_k-1) -(\mu-v_k) n_k \nonumber\\
&-& t_k (a_k a_{k+1}^+ + a_k^+ a_{k+1}) \Big]
\end{eqnarray}
where  $a_k^+$, $a_k$ and $n_k = a_k^+ a_k $ are respectively the boson creation, annihilation and number operators relevant to the site labeled $k$. As to the Hamiltonian parameters, $U>0$ accounts for on-site repulsion (proportional to the atomic scattering length), 
$\mu$ is the grand canonical chemical potential, 
$v_k$ is the local potential at site $k$ and $t_k$ is the hopping amplitude between nearest neighbour sites $k$ and $k+1$. According to the tight-binding-like approach of Ref.~\onlinecite{A:Jaksch}, on $\ell$-periodic optical superlattices
\begin{equation}
\label{E:parms}
t_k = t\, \tau_{{\rm mod}(k,\ell)+1}, \quad v_k =v \nu_{{\rm mod}(k,\ell)+1},
\end{equation}
where $\tau_h$ and $\nu_h$ (with $h=1,2,\ldots,\ell$) describe the modulation of the hopping amplitude and local potential within each cell, while $t$ and $v$ are scaling factors.  


The zero-temperature phase diagram of Hamiltonian (\ref{E:BHs}) can be characterized by the variation in the ground-state (free) energy produced by a finite variation
of the number of bosons in the system. In details, the system is either in an insulator or superfluid state  depending on whether such energy variation remains finite or vanishes in the thermodynamic limit $M \to \infty$. 
Note that, since the number operator $\sum_k n_k$ commutes with Hamiltonian (\ref{E:BHs}), the ground-state energy of the latter can be conveniently studied in the subspaces relevant to a fixed number of particles. Denoting $E(N)$ the ground-state energy of $H$ restricted to the $N$-boson subspace,  a critical incompressible domain exists whenever the equation  $E(f M) = E(f M\pm 1)$ yields two distinct solutions $\mu_- \neq \mu_+$ in the thermodynamic limit. In this case $\mu_\pm$ give the boundary of the insulator domain relevant to the critical filling $f=N/M$ in the $\mu/U-t/U$ 
phase diagram. Note that the compressibility of the system, $\kappa = \partial_\mu N$, vanishes within these domains.

On simple lattices, $\nu_h=\nu=0$, $\tau_h=\tau=1$, 
the presence of the energy gaps characterizing the incompressible insulator domains can be explained intuitively based on the so-called {\it atomic limit}, $t \to 0$, where the sites of the lattice can be regarded as independent ``atoms'' \cite{A:Fisher}. 
In such limit it is easy to prove that the free energy is always minimized by an integer filling.
More precisely $N=0$ for $\mu < 0$, while $N=q M$ for $q \leq \mu/U < q+1$, where $q \in \mathbb{N}$. Notice that the entire set of non commensurate  populations $q M < N<(q+1) M$ equivalently minimizes the free energy for the singular value of the chemical potential $\mu = q U$. 
The gain in kinetic energy produced by switching a small hopping amplitude on is sufficient to  make these configurations favourable on a narrow $\mu$ interval around $q U$, and the interval pertaining to the integer-filling configuration consequently shrinks by a small amount.  
Several numerical and analytical approaches (See Ref.~\onlinecite{A:Jain} and references therein for a review) show that the region of the $\mu/U-t/U$  enclosed by the boundaries $\mu_\pm$ has a lobe-like shape whose height decreases with increasing $f$.

In general, on $\ell$-periodic superlattices incompressible domains occur with critical rational fillings $f = N/M = q/\ell$, where $q\in {\mathbb N}$ \cite{A:Roth03,A:LobiMF,A:Santos}. This is quite easily  understood in the extreme  situation where the cells are drastically isolated from each other, i.e. when the hopping amplitudes between neighbouring cells are much smaller than those within the same cell \cite{A:Santos}. This is basically a {\it super-atomic limit}, the isolated cells being the super-atoms. In this situation the rational-filling $f=q/\ell$ of the system corresponds to an integer-filling  $f=q$ in terms of super-atoms.
A similar argument applies  when the cell features uneven local potentials $v_k$ \cite{A:Roth03,A:LobiMF}. Again, in the atomic limit $t \to 0$, the rational filling  
$f=q/\ell$ corresponds to an integer-filling 
$f=q$ for the energetically more favourable subset of atoms obtained extracting from each cell the ${\rm mod}(q,\ell)+1$ sites featuring the lowest local potentials.
In general, assuming that $\max(\{v_k\})-\min(\{v_k\}) \leq U$ and, without loss of generality, that $\min(\{v_k\}) =0$, the chemical potential interval $q < \mu/U< q+1$, $q\in {\mathbb N}$, is divided into $\ell$ parts, the $j$-th subinterval being the basis of the insulator domain relevant to the rational filling $f = q + j/\ell$. The points marking the boundaries of these intervals are directly related to the local potentials, $\mu = N U + v_j$, $j=1,\ldots,\ell$ \cite{A:LobiMF}. 
The condition $\max(\{v_k\}) > U$, discussed to some extent in Ref.~\onlinecite{A:LobiMF}, makes things more complex, and we do not address it here. 
Note that  whenever the local potentials $v_k$ are not all different from each other, some of the above intervals reduce to singular points, and no insulator domain seems to correspond to the relevant rational fillings.

In this paper we show that the zero-temperature phase diagram of a BH model on a 1D $\ell$-periodic SL features critical $q/\ell$ rational filling
insulator domains for every $q \in {\mathbb N}$. That is to say, an incompressible insulator domain exists also for those rational fillings that are predicted to be compressible in the atomic limit. We prove this by exploiting the exact mapping between the hard-core limit ($t/U\to 0$) of Hamiltonian (\ref{E:BHs}) and 
free spinless fermions in one dimension \cite{A:Girardeau}.  We furthermore investigate the boundary of these domains showing that they feature an unusual loophole shape. The presence of loophole insulator domains is qualitatively predicted by a {\it multiple-site} mean-field approach generalizing the one introduced in Ref.~\onlinecite{A:Sheshadri}. Quantitative results are obtained via quantum Monte Carlo (QMC) simulations. Furthermore, an analytic perturbative description of the boundaries for the half-filling insulator domain of the special case $\ell = 2$, $v_1=v_2 = 0$ is obtained which proves quite satisfactory.


\begin{figure}
\begin{center}
\includegraphics[width=8.5cm]{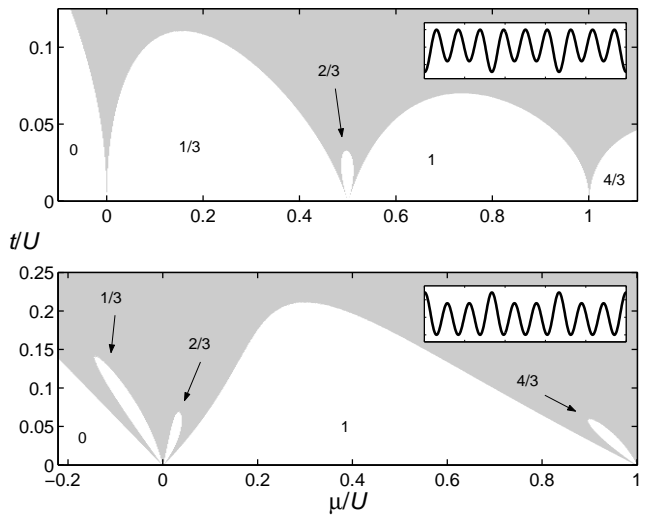}
\caption{\label{F:mfL} Zero-temperature phase diagrams for two simple $\ell~=~3$ superlattices (sketched in the insets). {\bf Upper panel}: $\tau_1 = \tau_2=\tau_3= 1$, $\nu_1=0$, $\nu_2 = \nu_3 = 0.5$, $v=U$. {\bf Lower panel}: $\tau_1 = \tau_2=1$, $\tau_3=0.3$, $\nu_1=\nu_2 = \nu_3 = 0$. }
\end{center}
\end{figure}

As we mention, the hard-core limit, $t/U \ll 1$, 
of 1D Hamiltonian  (\ref{E:BHs}) 
can be mapped exactly onto the tight-binding Hamiltonian describing spinless non-interacting fermions on the same ambient lattice \cite{A:Girardeau}, $H_{\rm F} = \sum_{k=1}^{M} [(\mu - v_k) c_k^+ c_k - t_k (c_k c_{k+1}^+ +  c_{k+1} c_k^+)] $, where $c_k$ is the fermionic annihilation operator and $\{c_k, c_h^+\}=\delta_{k h}$ \cite{N:HCB}. 
Since the fermions do not interact, the zero-temperature many-body ground state of $H_{\rm F}$ is simply obtained in terms of corresponding  single particle spectrum. As it is well-known, on a $\ell$-periodic SL, the latter features $\ell$ disjoint bands, each containing $M/\ell$ states \cite{N:slspec}. 
Hence, for critical rational fillings $q/\ell$ ($q=1,\ldots,\ell$) the fermionic system features an insulating ground-state which corresponds to an incompressible Mott state of the bosonic system. This proves that an insulating domain exists for any rational filling, including the ones that are not critical in the atomic limit. In particular, the latter feature a cusp rather than a wide basis for $t/U \to 0$. 
\begin{figure}
\begin{center}
\includegraphics[width=8.5cm]{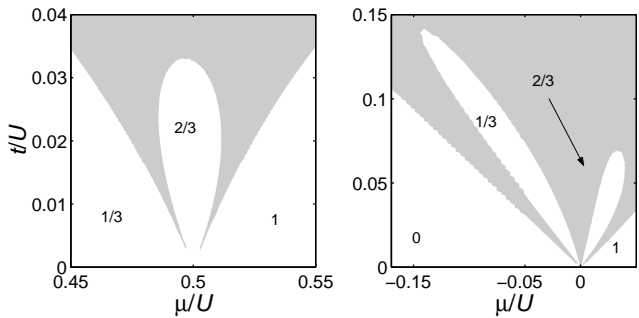}
\caption{\label{F:mfS} Left and right panel show a magnification of the first loophole insulator domains appearing in the upper and lower panels of Fig.~\ref{F:mfL}, respectively.}
\end{center}
\end{figure}

Qualitative information about the shape of these domains can be gained by considering a {\it multiple-site} mean-field approximation \cite{A:LoopLP} to Hamiltonian (\ref{E:BHs}). We recall that, following Ref.~\onlinecite{A:Sheshadri}, the latter can be recast as the sum of $M$ coupled single-site Hamiltonians by assuming that, for every $k$,
\begin{equation}
\label{E:MFA}
a_k a_{k+1}^+ = \langle a_k \rangle a_{k+1}^+ + a_k \langle a_{k+1}^+ \rangle  - 2 \langle a_k \rangle \langle a_{k+1}^+ \rangle,
\end{equation}
where $\langle \cdot \rangle$ denotes expectation value on the ground state. This approach gives a qualitative description of the fractional insulator lobes when the local potentials within a cell are different from each other \cite{A:LobiMF}, but fails to predict the existence of loophole insulator domains. 
If the non-trivial periodicity of the SL is taken into account by adopting approximation (\ref{E:MFA}) every $\ell$-th site, Hamiltonian   (\ref{E:BHs}) can be recast as the sum of $\ell$-site Hamiltonians. Note that the ensuing self-consistency problem is much more demanding than the usual single-site mean-field approach. Indeed, in general it requires the iterative diagonalization of a matrix whose size is $\sum_{q=0}^{Q} \frac{(q+\ell-1)!}{q! (\ell-1)!}$ where $Q$ provides a cutoff for the  infinite 
 Hilbert space of the problem \cite{N:num}, and must be sufficiently larger than the typical population for a given choice of the Hamiltonian parameters \cite{N:amf}.
We mention that a two-site mean-field approach is adopted in Ref.~\onlinecite{A:Jain} for the study of homogeneous lattices.

Figure~\ref{F:mfL} shows  zero-temperature phase-diagrams for two simple $\ell=3$ superlattices featuring insulator loophole domains, as evaluated by the multiple-site mean-field self-consistent approach. As usual, the insulator domains (white areas) are defined by the vanishing of the {\it superfluid order parameter}, $\langle a_k \rangle$ at every site $k$ of the SL. 
The relevant effective optical potentials are schematically depicted in the insets. Figure~\ref{F:mfS} shows a magnification of the loophole insulator domains appearing in Fig.~\ref{F:mfL}. 

A more precise approximation of the insulator domain borders can be achieved by means of perturbative approaches. On homogeneous lattices $E(f M)$ and $E(f M \pm 1)$ can be evaluated perturbatively, $t$ being the perturbative parameter \cite{A:Freericks2}. A reasonable effort allows to extend this approach  to superlattices, for the  insulator domains that are predicted by the atomic limit $t \to 0$. 
Analytic results  for the entire class of superlattices with constant local potential, where the atomic limit predicts integer filling domains only, are reported in Ref.~\onlinecite{A:scpe} .

Similarly to the {\it multiple-site} mean-field approach,
the loophole domains can be perturbatively studied by taking into account the periodicity of the SL, i.e. assuming that the perturbative parameter is the hopping term between adjacent cells. More precisely, if $\min(\{\tau_k\}) = \tau_\ell$, one can write
\begin{widetext}
\begin{equation}
t \sum_{k=1}^M \tau_k (a_{k \ell}^+ a_{k \ell+1}+ a_{k \ell} a_{k \ell+1}^+)
=  t \Big[ \tau_\ell \sum_{s=1}^{M/\ell}  (a_{s \ell}^+ a_{s \ell+1}+ a_{s \ell} a_{s \ell+1}^+) 
+ \sum_{s=0}^{M/\ell-1} \sum_{h=1}^{\ell-1} \tau_h (a_{h + s \ell}^+ a_{h+1 + s \ell}+ a_{h + s \ell} a_{h+1 + s \ell}^+ ) \Big]
\end{equation}
\end{widetext}
and consider $\tau_\ell$ as the perturbative parameter. The results are expected to be satisfactory for SLs very close to the {\it super-atomic limit}, i.e. when $\tau_\ell \ll \tau_k$, $\forall k \neq \ell$. 
This approach can be carried out analytically for the lowest loophole domains of some simple SLs. For instance, the first-order analytical result for the half-filling loophole insulator domain in the simple case
 $\ell = 2$, $v_1 = v_2$ is
\begin{eqnarray}
\label{E:mum}
\mu_- &=& (\tau_2 - \tau_1) t  \\
\label{E:mup}
\mu_+ &=& t \left[\tau_1 + \frac{\tau_2}{2} 
\frac{\left(4 t \tau_1 +U + R\right)^2}{16 (t \tau_1)^2 + U^2 + U R}\right]+\frac{U-R}{2} 
\end{eqnarray}
where $R = \sqrt{U^2 +16 (t \tau_1)^2}$.
\begin{figure}
\begin{center}
\includegraphics[width=8.5cm]{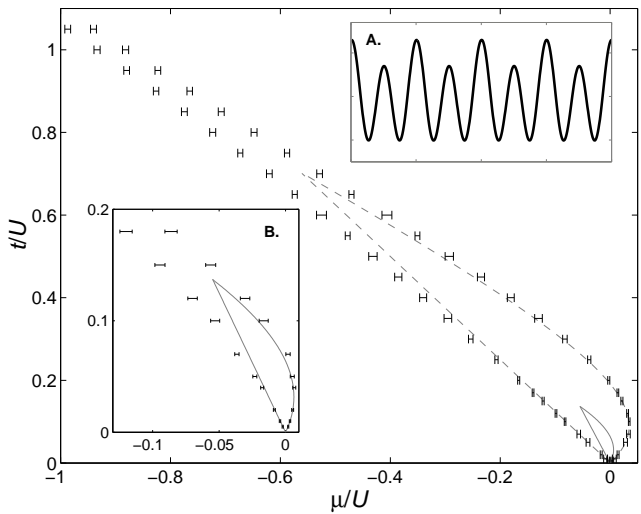}
\caption{\label{F:PEQMC} {\bf Inset A.:} Schematic representation of the optical potential characterizing the $\ell =2$ SL under concern. The relevant parameters are $\nu_1=\nu_2=0$ ($v=0$), $\tau_1 \neq \tau_2$.  {\bf Inset B.:}  Boundary of the half-filling loophole insulator for the parameter choice
$\tau_1=1$, $tau_2 = 0.6 $. Solid curves:  analytical result according to first order perturbative expansions, Esq.~(\ref{E:mum}) and (\ref{E:mup}); Data points: QMC simulations.  {\bf Main plot:}  Boundary of the half-filling loophole insulator for the parameter choice $\tau_1=1$, $\tau_2 = 0.2$ according to the perturbative expansion (dashed curves) and QMC simulations (data points). The perturbative result of inset B. is also shown for comparison (solid curves). The broadening of the QMC errorbars with increasing $t/U$ corresponds to the increase in the number of Fock states contributing to the ground state, which makes the estimate of the energy more noisy \cite{N:pQMC}. Furthermore, finite-size effects introduce an upper bound on the precision of the estimate at the tip of the domain. } 
\end{center}
\end{figure}
Fig.~\ref{F:PEQMC} shows the loophole domain borders as obtained by Eqs. (\ref{E:mum}) and (\ref{E:mup}) for the parameter choices $\tau_1 = 1$ and $\tau_2 = 0.2$ (solid curves) and $\tau_2 = 0.6$ (dashed curves).
The data points in Fig.~\ref{F:PEQMC} are the numerical QMC estimates of $\mu_\pm$ for a SL comprising  $M = 100$ sites. 
First of all we notice that these data points quantitatively confirm the loophole shape of the fractional incompressible domain under concern. On the other hand they show the perturbative results to be quite satisfactory also far from the {\it super-atomic limit}, $\tau_2\ll \tau_1$. 

The algorithm we adopted for the numerical evaluation of the data points in Fig.~\ref{F:PEQMC}
is the so-called population QMC, which basically amounts to a stochastic version of the {\it power method} for finding the maximal eigenvalue of a matrix \cite{N:pQMC}. 

In summary, in this paper we prove that the zero-temperature phase-diagram of a $\ell$-periodic 1D superlattice BH Hamiltonian features a $q/\ell$-filling incompressible insulator domain for every $q \in {\mathbb N}$. Furthermore we show that the insulator domains that are not predicted in the atomic limit feature an unusual loophole shape. It is worth remarking that such shape implies quite interesting reentrant behaviours. As it is clear from Fig.~\ref{F:mfS}, for some values of the chemical potential, the system rather unusually enters an insulating phase  as a result of an increase in kinetic energy. A similar behaviour is observed also in the first lobe of the 1D simple lattice \cite{A:Kuehner}. Even more unusually, two distinct insulator domain separated by a superfluid stretch can occur at the same $\mu/U$. The presence of the loophole insulator domains can be possibly detected as described in Ref.~\cite{A:Greiner}. Recently, an alternative scheme has been proposed, based on the measure of the Mott domain width via the laser probing of the particle-hole energy gap characterizing the insulating phase \cite{CM:Konabe}. This scheme seems suitable for detecting the reentrant character of the loophole domains. Indeed, as a function of $t/U$, the measured width is expected to feature a maximum for loophole domains, whereas it is simply non increasing for the usual lobes \cite{CM:Konabe}. 

We emphasize that the above considered 1D superlattices are quite realistic. Indeed  two superposing laser beams give rise to a 1D optical lattice whose period is determined by their crossing angle \cite{A:Morsch}. When two such lattices with commensurate period are superimposed, a 1D optical superlattice is obtained \cite{A:Peil}. This effective trapping potential can be used to fragment an elongated, cigar-shaped Bose-Einstein condensate, giving rise to a system described by Hamiltonian (\ref{E:BHs}). Ref.~\onlinecite{A:Peil} reports on the realization of a $\ell=3$ 1D optical superlattice where ultracold bosons are trapped. 
Note furthermore that the optical potentials in Fig.~\ref{F:mfL} basically differ by an overall sign. This means that in principle it is possible to switch between them by changing the sign of laser detuning.
We mention that the phase diagrams in Figs.~(\ref{F:mfL})-(\ref{F:PEQMC}) are obtained assuming that only $t$ is varied, for instance by varying the laser beam intensity $I$, while the remaining parameters in Eq.~(\ref{E:parms}) remain fixed. This is quite natural when there is no spatial modulation in the local potentials, and one can safely set $\nu_h = 0$. This not being the case, the linear dependence of $v$ on $I$ can be ignored to a first approximation, since $t\sim e^{-\gamma I}$. If this is not satisfactory, the desired effect can be obtained by adjusting both the laser intensity and the beam setup. 

The work of P.B. has been entirely supported by the
MURST project "Quantum Information and Quantum Computation
on Discrete Inhomogeneous Bosonic Systems." A.V.
also acknowledges partial financial support from the same
project.



\begin{thebibliography}{10}
\bibitem{N:num}
Recall that, when approximation~(\ref{E:MFA}) is adopted, the total number of
  bosons is no longer conserved, and one cannot consider one fixed-number
  subspace at a time.
\bibitem{N:amf}
An analytic description of the low-filling loophole domains is possible for
  some simple superlattices. We report about it elsewhere \cite{A:LoopLP}.
\end{thebibliography}

\end{document}